\documentclass[aps,preprint,superscriptaddress]{revtex4-1}

\usepackage{amsmath}    % need for subequations
\usepackage{graphicx}   % need for figures
\usepackage{color}      % use if color is used in text
\usepackage{subfigure}  % use for side-by-side figures
\usepackage{hyperref}   % use for hypertext links, including those to external documents and URLs
\usepackage{todonotes}
\usepackage{amssymb}
\usepackage[T1]{fontenc}
\usepackage[utf8]{inputenc}

\begin{document}

\title{Transverse resonance island buckets for synchrotron-radiation based electron time-of-flight spectroscopy}
\author{T. Arion}
\affiliation{Center for Free-Electron Laser Science / DESY, Notkestra\ss e 85, D-22607 Hamburg, Germany}
\author{W. Eberhardt}
\affiliation{Center for Free-Electron Laser Science / DESY, Notkestra\ss e 85, D-22607 Hamburg, Germany}
\affiliation{TU Berlin, Institute of Optics and Atomic Physics, Hardenbergstra\ss e 36, 10623 Berlin, Germany}
\author{J. Feikes}
\affiliation{Helmholtz-Zentrum Berlin, Albert-Einstein-Stra\ss e 15, 12489 Berlin, Germany}
\author{A. Gottwald}
\affiliation{Physikalisch-Technische Bundesanstalt, Abbestra\ss e 2-12, 10587 Berlin, Germany}
\author{P. Goslawski}
\affiliation{Helmholtz-Zentrum Berlin, Albert-Einstein-Stra\ss e 15, 12489 Berlin, Germany}
\author{A. Hoehl}
\affiliation{Physikalisch-Technische Bundesanstalt, Abbestra\ss e  2-12, 10587 Berlin, Germany}
\author{H. Kaser}
\affiliation{Physikalisch-Technische Bundesanstalt, Abbestra\ss e 2-12, 10587 Berlin, Germany}
\author{M. Kolbe}
\affiliation{Physikalisch-Technische Bundesanstalt, Abbestra\ss e 2-12, 10587 Berlin, Germany}
\author{J. Li}
\affiliation{Helmholtz-Zentrum Berlin, Albert-Einstein-Stra\ss e 15, 12489 Berlin, Germany}
\author{C. Lupulescu}
\affiliation{TU Berlin, Institute of Optics and Atomic Physics, Hardenbergstra\ss e 36, 10623 Berlin, Germany}
\author{M. Richter}
\email{Mathias.Richter@ptb.de}
\affiliation{Physikalisch-Technische Bundesanstalt, Abbestra\ss e 2-12, 10587 Berlin, Germany}
\author{M. Ries}
\email{Markus.Ries@helmholtz-berlin.de}
\affiliation{Helmholtz-Zentrum Berlin, Albert-Einstein-Stra\ss e 15, 12489 Berlin, Germany}
\author{F. Roth}
\email{Friedrich.Roth@physik.tu-freiberg.de}
\affiliation{TU Bergakademie Freiberg, Institute for Experimental Physics, Leipziger Stra\ss e 23, 09599 Freiberg, Germany}
\author{M. Ruprecht}
\affiliation{Helmholtz-Zentrum Berlin, Albert-Einstein-Stra\ss e 15, 12489 Berlin, Germany}
\author{T. Tydecks}
\affiliation{Helmholtz-Zentrum Berlin, Albert-Einstein-Stra\ss e 15, 12489 Berlin, Germany}
\author{G. W\"ustefeld}
\affiliation{Helmholtz-Zentrum Berlin, Albert-Einstein-Stra\ss e 15, 12489 Berlin, Germany}
\date{\today}

\begin{abstract}
At the Metrology Light Source (MLS), the compact electron storage ring of the Physikalisch-Technische Bundesanstalt (PTB) with a circumference of 48\,m, a specific operation mode with two stable closed orbits for stored electrons was realized by transverse resonance island buckets. One of these orbits is closing only after three turns. In combination with single-bunch operation, the new mode was applied for electron time-of-flight spectroscopy with an interval of the synchrotron radiation pulses which is three times the revolution period at the MLS of 160\,ns. The achievement is of significant importance for PTB's future programs of angular-resolved electron spectroscopy with synchrotron radiation and similar projects at other compact electron storage rings. Moreover, the applied scheme for orbit and source spot selection via optical imaging at the insertion device beamline of the MLS and may be relevant for the BESSY VSR project of the Helmholtz-Zentrum Berlin. 
\end{abstract}

\maketitle

Current developments of accelerator-based (soft) X-ray sources for materials research refer to two main directions: (a) single pass free electron lasers (FELs) based on self-amplified spontaneous emission (SASE) \cite{Ackermann2007,Mitri2011,Emma2010,Shintake2008,Pile2011}  or different seeding concepts \cite{Allaria2012,Schneidmiller2017} and (b) diffraction-limited storage rings (DLSRs) \cite{Tavares2014}. FELs provide for single-user experiments peak brightness femtosecond pulses at repetition rates in the kHz to MHz regime to study ultra-fast dynamical and non-linear processes. DLSRs are optimized, on the other hand, for diffraction limited high average brightness synchrotron radiation (SR), i.\,e., for achieving the ultimate lateral spot size on a sample and, e.\,g, spectro-microscopy applications at pulse durations ranging from some 10\,ps to more than 200\,ps. The advantage of DLSRs and generally storage rings is the possibility of multi-user operation at repetition rates beyond 1\,MHz. Here, the samples can be interrogated multiple times in the linear response regime enabling to make 'movies' of processes. While many storage ring facilities around the world are planning upgrades to achieve the diffraction limited photon beam, the BESSY VSR project at the Helmholtz-Zentrum Berlin (HZB)  \cite{Jankowiak2016} follows an alternative concept. This upgrade of the electron storage ring BESSY II in Berlin-Adlershof is aimed at providing short photon pulses down to 2\,ps (rms) in parallel with multi-user standard (\textgreater 15\,ps) operation and even down to 400\,fs (rms) in a special operation mode.

\par

BESSY VSR is a novel approach to create photon pulses of different length and period circulating simultaneously for all SR beamlines by using a pair of superconducting cavities shaping the longitudinal phase space. Pulse-picking schemes at each beamline enable each individual user to freely switch, e.\,g., between high average photon flux for X-ray spectroscopy, microscopy and scattering and picosecond pulses for dynamic studies. Preparatory work on the BESSY VSR concept has, however, been performed at the Metrology Light Source (MLS) of the Physikalisch-Technische Bundesanstalt (PTB), Germany's national metrology institute \cite{Gottwald2012}. The MLS represents a compact 630\,MeV electron storage ring with a circumference of 48\,m in the immediate vicinity of BESSY II and is dedicated to SR based metrology. Also operated by HZB, the MLS is optimized regarding its regular performance (electron beam current and lifetime) as well as for special operations with variable electron energies and bunch lengths. By design, it is equipped with additional families of sextupole and octupole magnets, and therefore is ideally suited to investigate nonlinear beam dynamics \cite{Gottwald2012,Feikes2011,Ries2014,Ries2015}.

In particular, the population of so-called transverse resonance island buckets (TRIBs) has successfully been realized at the MLS \cite{Ries2015}, i\,.e., the simultaneous operation of two orbits. It offers the possibility to store two distinct transversely displaced electron beams, e.\,g., homogeneous fill and pseudo single bunch, separated in angle and space. Thus, apart from pulse-picking schemes, limited in some ways, the different bunch modes may be simply selected by means of optical imaging at individual SR beamlines, respectively, which may be relevant for BESSY VSR. Optical selection schemes have already been realized at fs-slicing sources at the Advanced Light Source in Berkeley \cite{Schoenlein2000} and at BESSY II \cite{Holldack2006,Holldack2014}. For compact storage rings like the MLS, TRIBs operation provides, however, a further promising aspect: a single bunch trapped in an island in the first turn will be observed in another island in the second turn and so forth. In case of a 3rd order resonance it will come back to the first island bucket again at the 4th turn. This property may be exploited to generate photon pulses with sub-revolution repetition rates. The latter allows, even at smaller rings with high revolution frequencies, electron time-of-flight (eTOF) spectroscopy to be performed. eTOF spectroscopy is usually applied in the single-bunch mode of a storage ring and based on the evaluation of electron kinetic energies from the interval between a SR pulse, triggered by the bunch-clock signal, and the electron detection. The corresponding start-stop measurements are, however, significantly disturbed if the electron time-of-flights of typically up to about 400\,ns exceed the photon pulse interval. 

\begin{figure}[h]
\includegraphics[width=.8\linewidth]{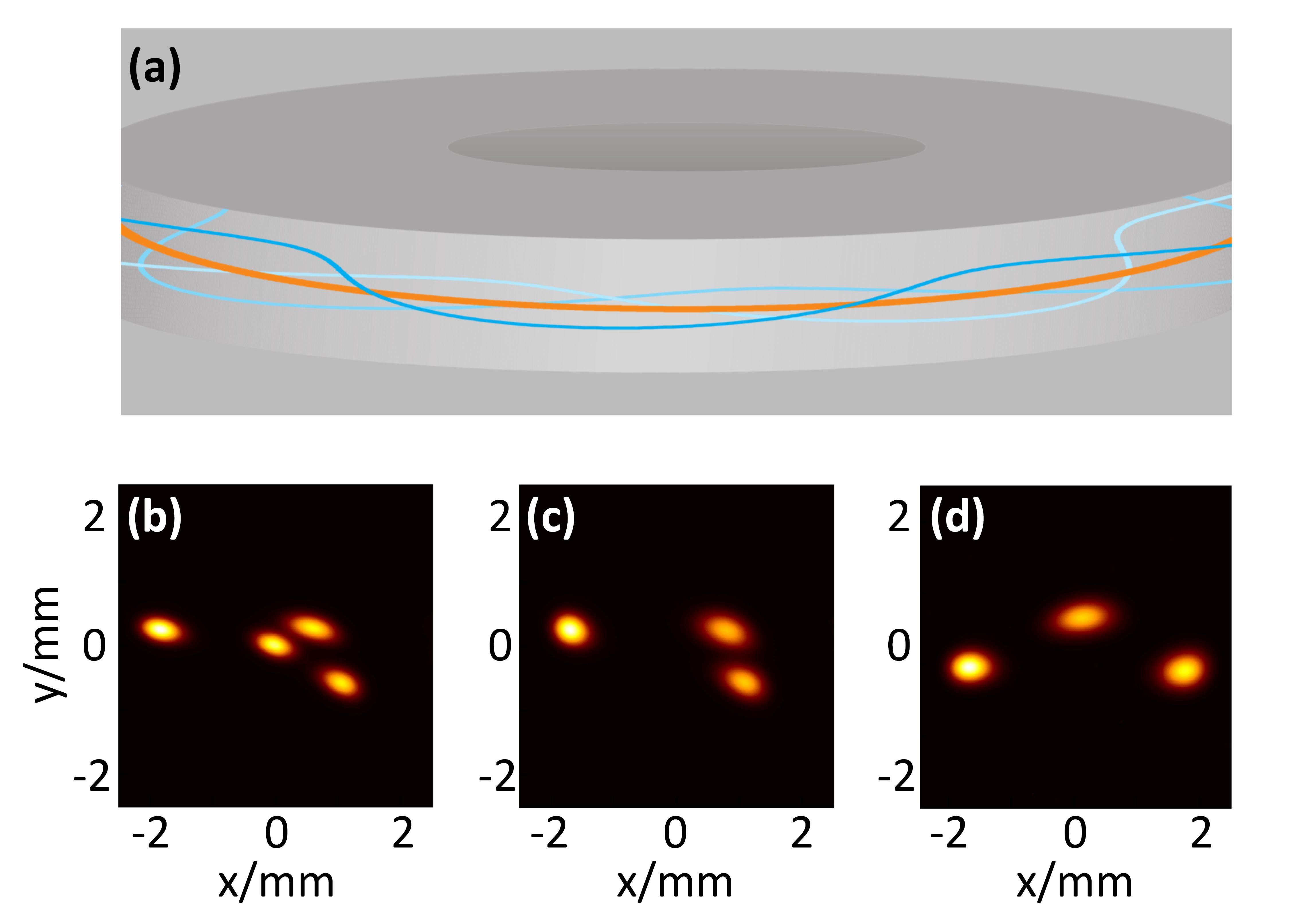}
\caption{(a): Sketch of the TRIBs orbit at the MLS closing after three turns (different shades of blue), twisting around the standard orbit (orange), i.\,e., being transversely separated in space and angle. (b): Cross section of the electron beam during TRIBs orbit single bunch operation of the MLS measured by using a source spot imaging system. (c): The same as (b), however, with empty regular standard orbit in the centre (at x = y = 0 position). (d) The same as (c), however, by using another source spot imaging system at a different position along the storage ring \cite{Ries2015}.}
\label{fig1}
\end{figure}

\par

In the present proof-of-principle study, we demonstrate both: (a) orbit selection by means of optical imaging at the insertion device beamline of the MLS during TRIBs operation and (b) its application for eTOF spectroscopy which adds a new quality for users of compact storage rings. In Figure \ref{fig1}\,(a), the standard orbit (orange) is illustrated in comparison to a specific TRIBs orbit closing after three turns (different shades of blue) which may be stored simultaneously \cite{Ries2015}. In general, both orbits are transversely separated in space and/or angle. Thus, at certain cuts perpendicular to the beam direction, one obtains simultaneously four source spots, one of the standard orbit and three of the threefold TRIBs orbit as shown in  Fig.\ref{fig1}\,(b). The image in Fig.\ref{fig1}\,(c) was obtained while the regular standard orbit was not populated so that only the three source spots of the threefold TRIBs orbit are visible and Fig.\ref{fig1}\,(d) by means of another source spot imaging system at a different position along the storage ring and a different spatial source spot distribution.

\begin{figure}[h]
\includegraphics[width=.55\linewidth]{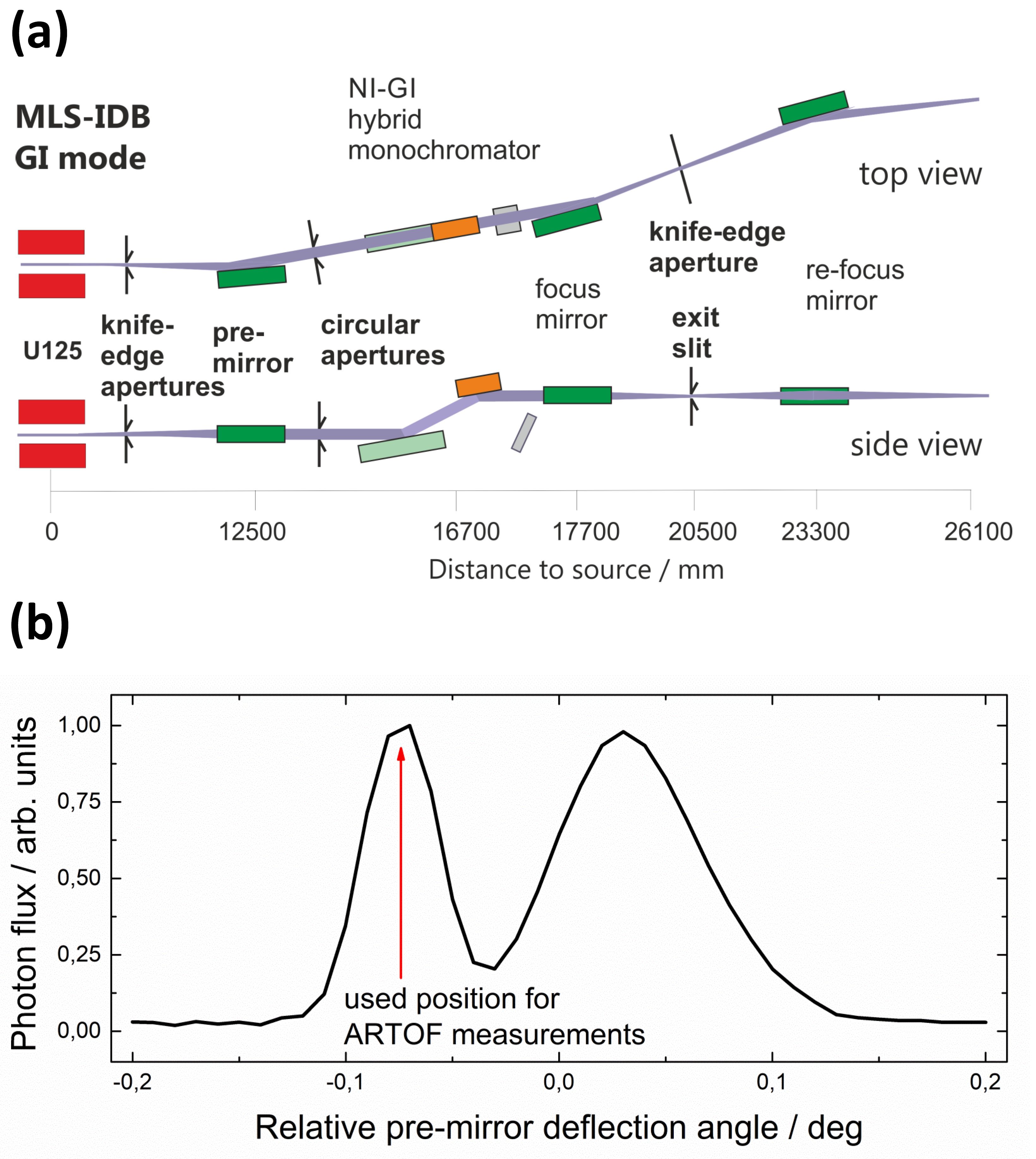}
\caption{(a): Sketch of the insertion device beamline (IDB) at the MLS combining a gracing incidence (GI) and a normal incidence (NI) monochromator. The source spot is imaged onto the plane of the exit slit. (b): Photon flux behind the exit slit as a function of the horizontal pre-mirror deflection angle relative to the standard position (set to 0).}
\label{fig2}
\end{figure}

\begin{figure}[h]
\includegraphics[width=.5\linewidth]{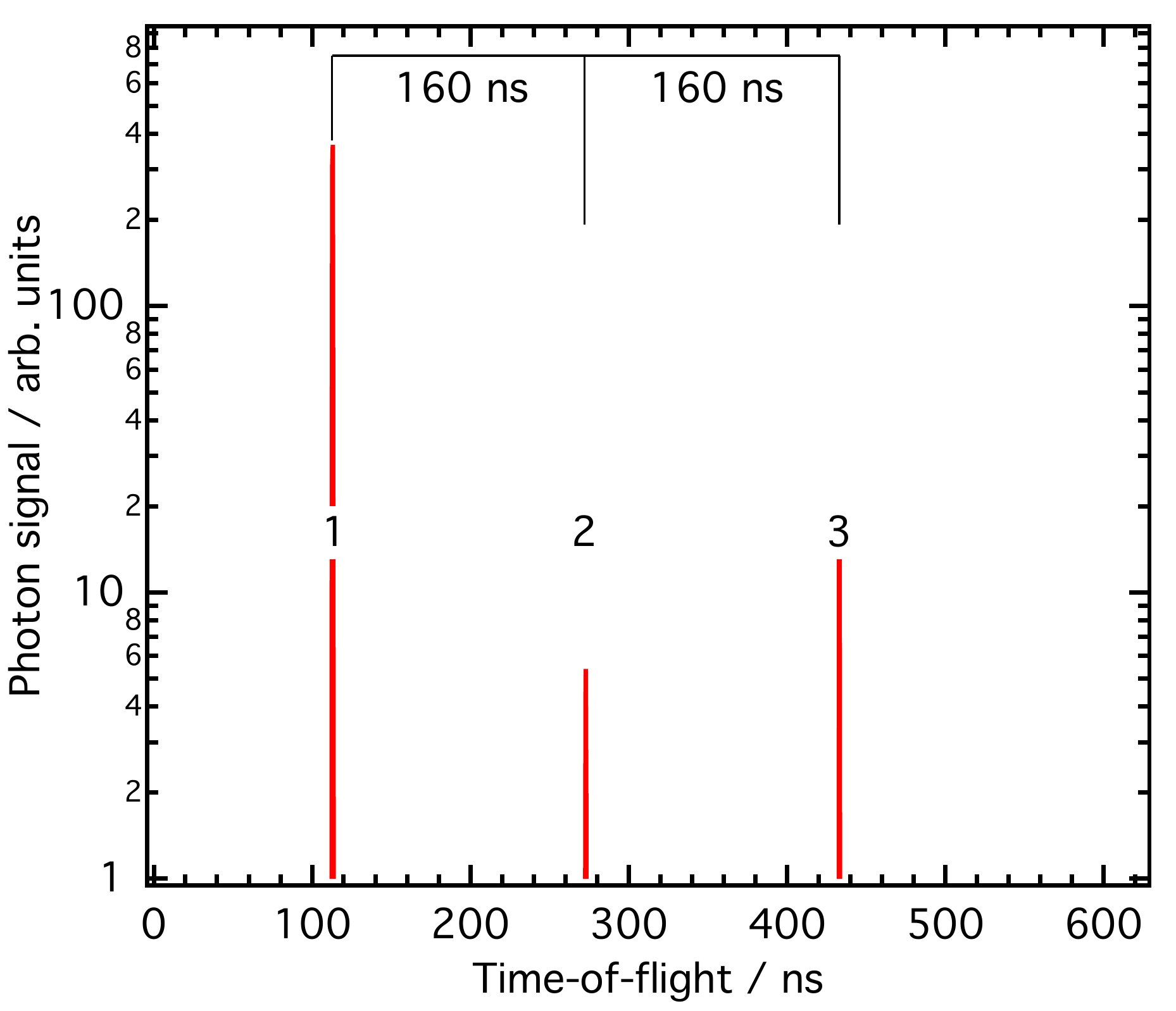}
\caption{Time-of-flight spectrum recorded with the ARTOF spectrometer using a high bias voltage on the detector mesh in order to reject electrons and accept photons scattered from the sample only.}
\label{fig3}
\end{figure}

In the configuration described, i.\,e., in the threefold TRIBs orbit mode with empty central orbit, the present measurements were performed at the insertion device beamline (IDB) of the MLS which is sketched in Fig.\ref{fig2}\,(a). IDB utilizes photons from the MLS undulator with 125\,mm period length (U125) in the photon energy range from 1.5 eV to 285 eV by means of a normal incidence grazing incidence (NI-GI) hybrid monochromator with the source spot being imaged onto the exit slit plane. In order to select one of the three source spots of the TRIBs orbit and suppress the other two as much as possible, the exit slit and different beamline apertures were used in combination with a proper adjustment of the pre-mirror. The spatial source spot discrimination achieved at IDB is demonstrated in Fig.\ref{fig2}\,(b) which shows the relative photon flux behind the exit slit (at the photon energy of 31\,eV) as measured with the help of a gas photoionization cell while scanning the horizontal deflection angle of the pre-mirror. Since the spatial distribution of the three TRIBs spots at U125 corresponds almost to the image in Fig.\ref{fig1}\,(c), i.\,e., a good separation of one single spot from the other two in horizontal (x) direction, the two maxima in Fig.\ref{fig2}\,(b) can be attributed accordingly, i.\,e., the single spot relates to the left peak and the other two to the broader peak at the right.

At the position marked in Fig.\ref{fig2}\,(b), i.\,e., with photons basically from one single source spot of the threefold TRIBs orbit filled with one single electron bunch, measurements were performed with a linear eTOF spectrometer of the Scienta ARTOF 10k (Scienta Omicron, Uppsala, Sweden) type \cite{Ovsyannikov2013}. By means of the IDB re-focus mirror Fig.\ref{fig2}\,(a)), the photon beam was  guided onto an Au (111) single crystal sample within a vacuum chamber equipped with the ARTOF and a second electron spectrometer, namely a electrostatic hemispherical electron energy analyser of the Scienta R4000 (Scienta Omicron, Uppsala, Sweden) type \cite{OHRWALL2011}. The two spectrometers were arranged in an orthogonal geometry and are facing the same interaction region on the sample. The sample surface was cleaned by performing repeatedly cycles of Ar ion sputtering and annealing at 650 degrees Celsius. The background pressure during the measurements was in the order of 7x10$^{-8}$\,Pa.  

\par

Figure\,\ref{fig3} shows an ARTOF spectrum taken at the photon energy of 44\,eV employing, however, a so high bias voltage on the mesh in front of the ARTOF multi-channel plate (MCP) detector that electrons were rejected. Thus, only photons scattered from the sample surface could reach the MCP. From the period of revolution at the MLS, one would expect, in the single bunch mode, every 160\,ns a photon pulse which is usually too short for eTOF spectroscopy. However, the second and third peak in Fig. \,\ref{fig3} are strongly suppressed which reflects the almost complete discrimination of one single orbit (\#1) at IDB in the TRIBs mode at the MLS with 5 \% contribution from the other two orbits (\#2 and \#3) only. This contribution can be further reduced at the expense of photon flux.

\begin{figure}[t]
\includegraphics[width=.5\linewidth]{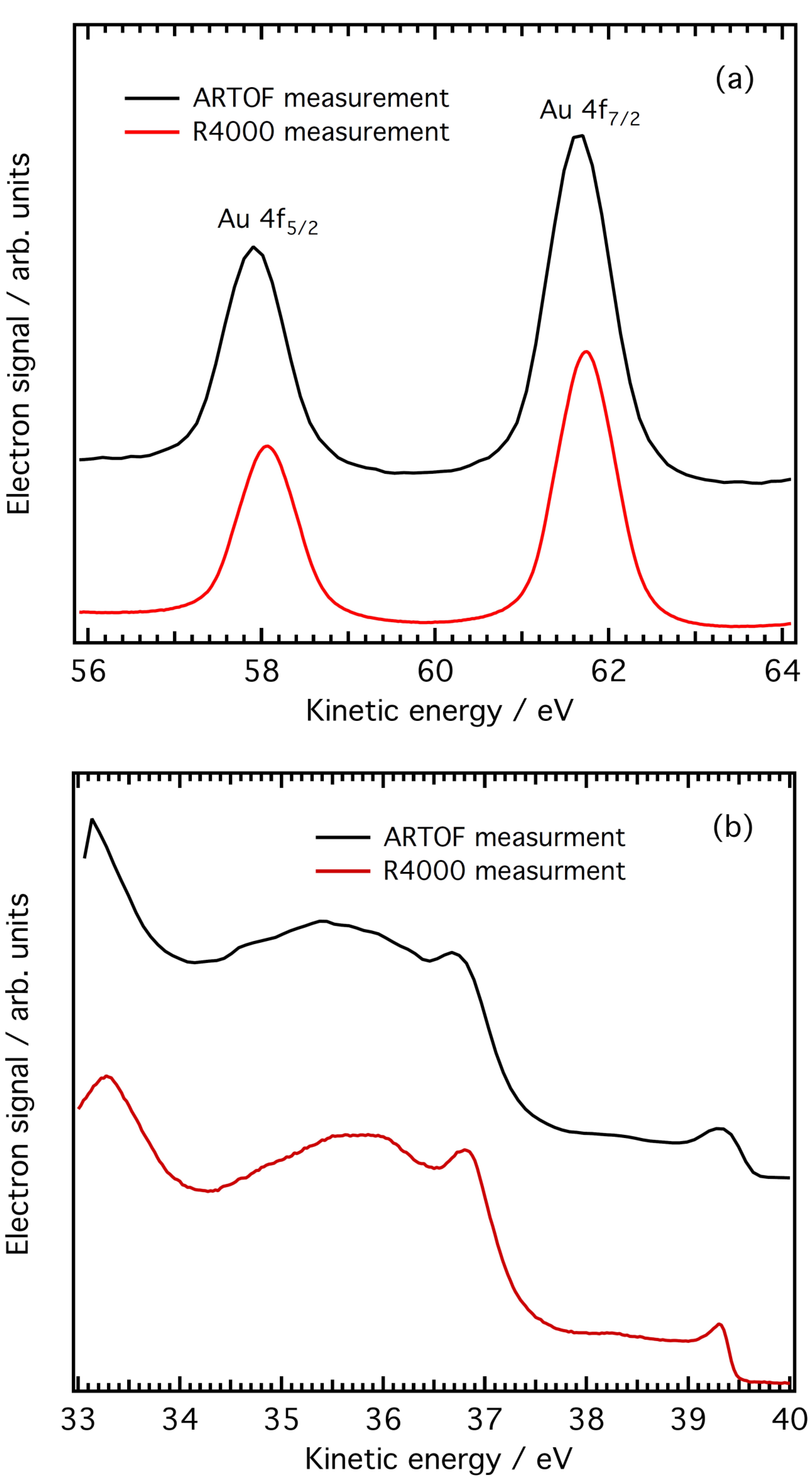}
\caption{Comparison of (a) the Au\,4f electron energy spectra (at 150\,eV photon energy) and (b) Au valence band electron energy spectra (at 44\,eV photon energy) as recorded using the ARTOF spectrometer in the TRIBs orbit single bunch operation mode of the MLS at a ring current of below 1\,mA (black upper curves) and the R4000 analyser in the regular standard orbit multi-bunch operation mode of the MLS at a ring current beyond 100\,mA (red lower curves). }
\label{fig4}
\end{figure}

Thus, an operational mode at the IDB/MLS was realized with a photon pulse interval of 3\,x\,160\,ns = 480\,ns, i.\,e., sufficient for eTOF spectroscopy. At a ring current of below 1\,mA, it was used to measure eTOF spectra of the Au (111) sample using the ARTOF spectrometer. Figure \ref{fig4}\,(a) shows the spectrum of the Au 4f electrons obtained at the photon energy of 150\,eV and Fig.\,\ref{fig4}\,(b) of the Au valence band obtained at 44 eV (black upper curves), both compared, respectively, to results obtained with the R4000 analyser (red lower curves). The latter measurements were performed one day later, without breaking of the vacuum or manipulating the sample, in the regular standard orbit multi-bunch operation mode of the MLS at a ring current beyond 100\,mA, however at slightly reduced transmission and increased resolution of the IDB monochromator. Although the energy resolution of the ARTOF spectra is, therefore, a bit lower, the equivalence of the data sets obtained at comparable accumulation times is obvious, respectively. Taking into account the lower ring current and the at least 10 times lower photon flux during the ARTOF measurements demonstrates the tremendous capability of eTOF spectrometers, i.\,e., by far higher responsivity due to the simultaneous detection of the complete electron energy spectrum compared to electrostatic analysers.

The 60 eV electrons of Fig.\,\ref{fig4}\,(a)  and the 35\,eV electrons of Fig.\,\ref{fig4}\,(b) refer to a time-of-flight within the ARTOF in the order of 200\,ns to 300\,ns, respectively, i.\,e., longer than the single bunch period of revolution at the MLS of 160\,ns but significantly shorter than the photon pulse interval of 480\,ns of the new operational mode. Hence, selecting a single orbit by optical imaging of a threefold TRIBs orbit makes eTOF spectroscopy feasible at the MLS or other compact SR sources with a circumference shorter than 100\,m.

\par

In conclusion, two-orbit operation was realized at PTB's MLS by transverse resonance island buckets. Source spot selection was successfully demonstrated by means of optical imaging at the MLS insertion device beamline. The scheme might be of relevance for the upcoming BESSY VSR project in combination with SR beamline upgrades further optimized for multiple orbit and source spot discrimination. Moreover, a specific second orbit closing after three turns, which is based on transverse resonance island buckets, was tested. A photon pulse interval three times longer than the single bunch period of revolution qualifies the MLS for the application of highly efficient electron time-of-flight spectroscopy which is of significant importance for PTB's future programs of angular-resolved electron spectroscopy with synchrotron radiation together with external partners  \cite{Roth2015,Weiss2015,ROTH20162,Darlatt2016,Luftner2017,ROTH2018} and similar projects at compact storage rings in general.

% \bibliography{Island_Bucket}
%\bibliographystyle{apsrev4-1}

%merlin.mbs apsrev4-1.bst 2010-07-25 4.21a (PWD, AO, DPC) hacked
%Control: key (0)
%Control: author (72) initials jnrlst
%Control: editor formatted (1) identically to author
%Control: production of article title (-1) disabled
%Control: page (0) single
%Control: year (1) truncated
%Control: production of eprint (0) enabled
%

\end{document}